\documentclass[conference]{IEEEtran}
\IEEEoverridecommandlockouts

\usepackage{cite}
\usepackage{amsmath,amssymb,amsfonts}
\usepackage{graphicx}
\usepackage{booktabs}
\usepackage{multirow}
\usepackage{stfloats}
\usepackage{placeins}
\usepackage{url}

\begin{document}

\title{scVGAE: A ZINB-Based Variational Graph Autoencoder for Single-Cell RNA-Seq Imputation}

\author{\IEEEauthorblockN{Yoshitaka Inoue}
\IEEEauthorblockA{\textit{Department of Computer Science and Engineering} \\
\textit{University of Minnesota}\\
Minneapolis, MN, USA}}

\maketitle

\begin{abstract}
Single-cell RNA sequencing (scRNA-seq) provides high-resolution measurements of cellular heterogeneity, but sparsity and technical zeros can obscure biological structure and complicate downstream analysis. We present scVGAE, a variational graph autoencoder for scRNA-seq imputation that integrates cell-cell graph propagation, a zero-inflated negative binomial (ZINB) likelihood, and direct expression reconstruction. scVGAE constructs a scalable cell graph using principal component analysis (PCA) followed by $k$-nearest neighbors, and encodes each cell into the parameters of a Gaussian latent distribution using graph convolutional networks (GCNs). A low-dimensional latent representation is obtained through stochastic reparameterization and is decoded both into gene-wise ZINB parameters and into a reconstructed expression matrix. Training jointly optimizes ZINB negative log-likelihood, mean-squared reconstruction error, and Kullback--Leibler divergence regularization. 

We evaluate scVGAE on 14 real-world scRNA-seq datasets against the original expression data and five established imputation methods: MAGIC, ALRA, DeepImpute, DCA, and GNNImpute. scVGAE achieves the highest mean Adjusted Rand Index (ARI) of 0.4681 and the second-highest mean Adjusted Mutual Information (AMI) of 0.5729 across the 14 datasets. These results demonstrate that a compact variational graph representation can preserve cell-class structure competitively across heterogeneous datasets while simultaneously producing an imputed expression matrix.
\end{abstract}


\section{Introduction}

Single-cell RNA sequencing (scRNA-seq) has become an important
technology for characterizing cellular heterogeneity and identifying
cell populations at single-cell resolution~\cite{luecken2019current,heath2016single}.
It is also increasingly used in broader biomedical and drug-discovery
settings, where cell-type-specific expression profiles can help
connect molecular variation to disease-relevant cellular states
~\cite{inoue2026drug}. A persistent challenge, however, is
that scRNA-seq expression matrices are sparse and contain many zero
or weakly observed entries. Some zeros reflect biological absence of
expression, whereas others arise from limited sampling and technical
effects. This sparsity can reduce the stability of visualization,
clustering, and other downstream analyses.

A range of imputation and denoising methods has been proposed.
MAGIC~\cite{van2018recovering} diffuses information over a cell-cell
affinity graph, while ALRA~\cite{linderman2022zero} uses low-rank
approximation. DeepImpute~\cite{arisdakessian2019deepimpute} predicts
gene expression using neural networks, whereas
DCA~\cite{eraslan2019single} combines an autoencoder with a
count-distribution objective. Graph-based methods explicitly exploit
structured relationships among cells and genes;
GNNImpute~\cite{xu2021efficient} uses graph neural networks for
expression recovery, while BiGCN~\cite{inoue2024bigcn} leverages both
cell-cell and gene-gene similarities through bi-graph convolutional
networks for single-cell transcriptome imputation.

Graph neural networks (GNNs)~\cite{zhou2020graph} are particularly
well suited to scRNA-seq because cells can be represented as nodes
connected according to transcriptional similarity. However,
deterministic graph representations provide only a single latent
embedding for each cell and do not explicitly model uncertainty in
the latent state. A variational formulation instead represents each
cell through a latent posterior distribution, enabling stochastic
reparameterization and regularization toward a prior. This provides a
probabilistic graph representation that can be combined naturally with
count-aware reconstruction objectives for sparse scRNA-seq data.

We present scVGAE, a ZINB-based variational graph autoencoder for
single-cell RNA-seq imputation. scVGAE combines graph-convolutional
message passing with a Gaussian latent posterior, stochastic
reparameterization, and KL-divergence regularization. The latent
representation is decoded through both gene-wise ZINB parameter heads
and a separate expression decoder. To improve scalability, the model
constructs a PCA-based $k$-nearest-neighbor cell graph and performs
graph propagation in a low-dimensional hidden space.

The main contributions are:
(1) a graph-convolutional variational encoder that parameterizes a
Gaussian posterior for each cell;
(2) joint ZINB and expression reconstruction objectives for modeling
sparse scRNA-seq measurements;
(3) scalable PCA-based $k$-nearest-neighbor graph construction; and
(4) evaluation across 14 benchmark datasets against six reference and
imputation approaches.

\section{Related Work}
\subsection{Statistical and Affinity-Based Imputation}
Statistical approaches exploit low-rank, distributional, or local similarity structure in sparse expression matrices. ALRA~\cite{linderman2022zero} uses low-rank approximation while preserving observed zeros, whereas SDImpute~\cite{qi2021sdimpute} and scISR~\cite{tran2022novel} estimate missing expression using statistical structure at the cell or gene level. Affinity-based approaches instead propagate information among transcriptionally similar cells or genes. MAGIC~\cite{van2018recovering}, scImpute~\cite{li2018accurate}, and DrImpute~\cite{gong2018drimpute} use neighborhood or similarity information to recover weakly observed expression values. Although these approaches can effectively reduce sparsity, they do not explicitly learn graph-aware probabilistic latent representations.

\subsection{Deep Learning and Graph-Based Methods}
Deep-learning methods learn nonlinear mappings from sparse expression matrices to denoised or reconstructed representations. DeepImpute~\cite{arisdakessian2019deepimpute} predicts target genes from correlated genes, whereas DCA~\cite{eraslan2019single} combines a deep autoencoder with a count-distribution objective. Graph-based methods additionally exploit structured relationships among cells, genes, or both. Graph convolutional networks (GCNs) provide a widely used normalized message-passing formulation~\cite{kipf2017semi}. scGNN2.0~\cite{gu2022scgnn} integrates feature learning, graph construction, and clustering, while GNNImpute~\cite{xu2021efficient} uses graph neural networks for expression reconstruction. BiGCN~\cite{inoue2024bigcn} extends graph-based imputation by jointly modeling cell-cell and gene-gene similarities using bi-graph convolutional networks. scTAG~\cite{yu2022zinb} combines topology-adaptive graph convolution with a ZINB objective for single-cell clustering. These methods demonstrate the utility of graph structure for single-cell analysis, but most rely on deterministic latent representations.

\subsection{Variational Graph Representation}
Variational autoencoders (VAEs) learn probabilistic latent representations by combining stochastic reparameterization with KL-divergence regularization~\cite{kingma2013auto}. Variational graph autoencoders (VGAEs) extend this principle to graph-structured data by using graph convolutions to parameterize a latent posterior and, in the classical formulation, reconstruct graph connectivity ~\cite{kipf2016variational}. scVGAE adopts this probabilistic graph-encoding framework for single-cell expression imputation. Rather than reconstructing adjacency, its graph encoder parameterizes a Gaussian latent representation for each cell, and the sampled latent variables are decoded into both ZINB parameters and reconstructed gene expression. This formulation combines graph-aware representation learning with distribution-aware expression modeling.

\section{Methods}
\begin{figure*}[!t]
    \centering
    \includegraphics[width=\textwidth]{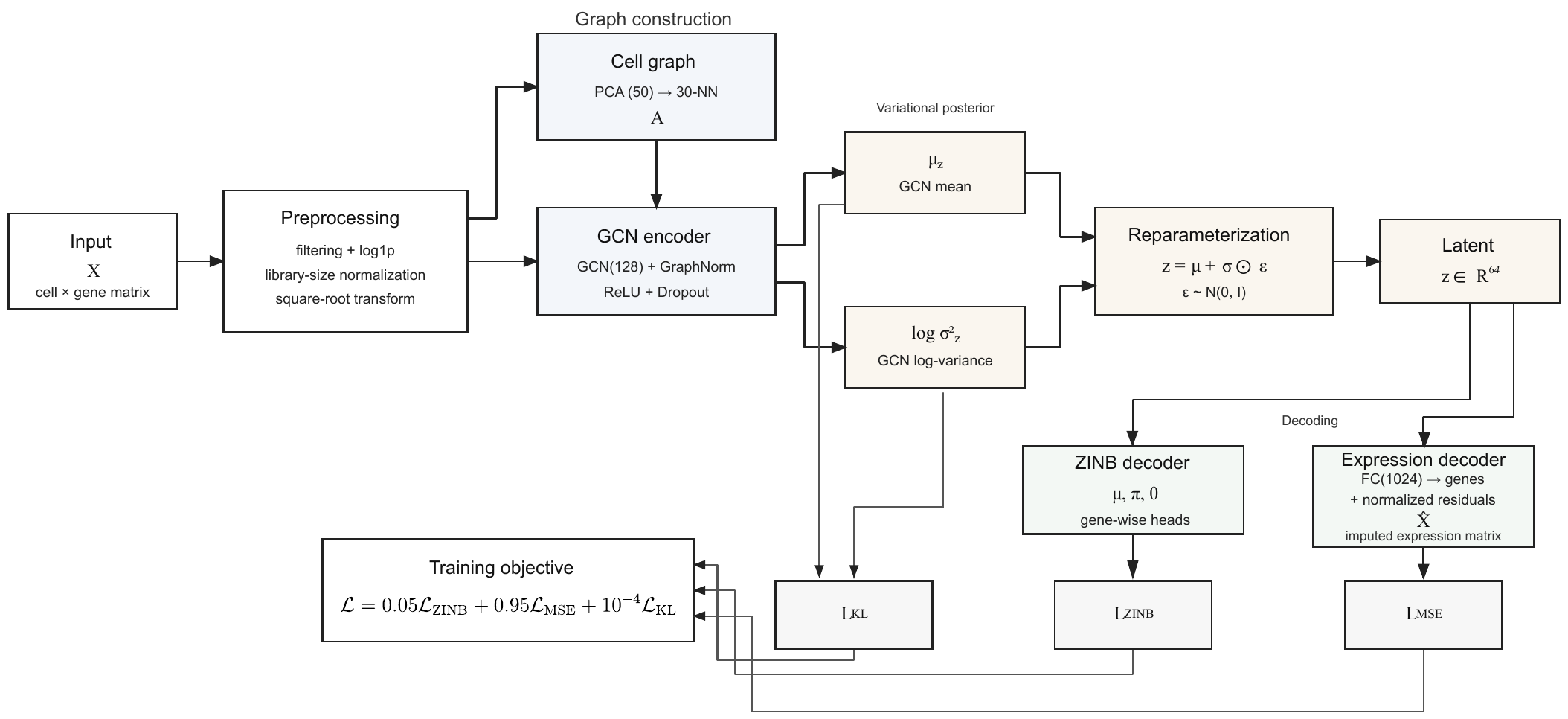}
    \caption{Architecture of scVGAE. The preprocessed cell-by-gene expression matrix is projected by PCA to construct a symmetric 30-nearest-neighbor cell graph. Graph convolutional layers encode each cell into the parameters of a Gaussian posterior, from which a 64-dimensional latent representation is obtained by reparameterization. The latent representation is decoded through gene-wise ZINB parameter heads and a separate expression decoder. Training jointly optimizes the ZINB negative log-likelihood, mean-squared expression reconstruction loss, and KL-divergence regularization.}
    \label{fig:architecture}
\end{figure*}

Fig.~\ref{fig:architecture} summarizes the scVGAE workflow, including graph construction, variational graph encoding, reparameterization, the ZINB and expression decoders, and the joint training objective.

\subsection{Preprocessing}
Let $X\in\mathbb{R}^{n\times m}$ denote the cell-by-gene expression matrix. We retain genes with non-zero expression in more than 5\% of cells and cells with non-zero expression in more than 5\% of genes. The retained counts are transformed as
\begin{equation}
X^{(1)}=\log(1+X).
\end{equation}
For each cell, the log-transformed profile is library-size normalized to a total of 10,000 and then square-root transformed. The resulting matrix is used as the model input and as the reconstruction target.

\subsection{Cell Graph Construction}
Dense all-pairs similarity graphs scale quadratically with the number of cells. To construct a bounded-degree graph, we first project the preprocessed expression matrix to a PCA representation with at most 50 components. Euclidean $k$-nearest neighbors are then computed with $k=30$. Directed neighbor relations are symmetrized and duplicate edges are removed. The resulting graph is represented by an edge set $E$ and supplied to graph-convolutional layers following the normalized GCN propagation principle~\cite{kipf2017semi}.

Self-loops and symmetric degree normalization are handled by \texttt{GCNConv}. For adjacency matrix $A$, the propagation operator is
\begin{equation}
\widetilde{A}=\widetilde{D}^{-1/2}(A+I)\widetilde{D}^{-1/2},
\end{equation}
where $\widetilde{D}$ is the degree matrix of $A+I$.

\subsection{Variational Graph Encoder}
Following the VAE framework~\cite{kingma2013auto} and the variational graph-encoding principle of VGAE~\cite{kipf2016variational}, scVGAE parameterizes an approximate Gaussian posterior for each cell using graph convolutions. Unlike classical VGAE, whose decoder reconstructs graph connectivity, scVGAE uses the sampled graph-aware latent representation to reconstruct gene-expression quantities.

The first graph-convolutional layer maps the gene-dimensional input to a 128-dimensional hidden representation and applies GraphNorm~\cite{cai2021graphnorm}:
\begin{equation}
H=\operatorname{Dropout}_{0.2}\!\left(\operatorname{ReLU}\left(\operatorname{GraphNorm}(\operatorname{GCN}_{1}(X,E))\right)\right).
\end{equation}
Two separate graph-convolutional layers parameterize the approximate posterior distribution:
\begin{equation}
\boldsymbol{\mu}_z=\operatorname{GCN}_{\mu}(H,E),\quad
\log\boldsymbol{\sigma}_z^2=\operatorname{clip}(\operatorname{GCN}_{\log\sigma^2}(H,E),-10,10).
\end{equation}
Both posterior parameters have latent dimension $d=64$. During training, a latent sample is obtained through the reparameterization trick,
\begin{equation}
Z=\boldsymbol{\mu}_z+\boldsymbol{\epsilon}\odot\exp\left(\frac{1}{2}\log\boldsymbol{\sigma}_z^2\right),
\quad \boldsymbol{\epsilon}\sim\mathcal{N}(0,I).
\end{equation}
At inference time, the posterior mean $\boldsymbol{\mu}_z$ is used as the latent representation. The KL-divergence term is
\begin{equation}
L_{KL}=-\frac{1}{2n}\sum_{i=1}^{n}\sum_{r=1}^{d}
\left(1+\log\sigma_{ir}^{2}-\mu_{ir}^{2}-\sigma_{ir}^{2}\right).
\label{eq:kl}
\end{equation}

\subsection{ZINB Decoder}
The latent representation is projected to per-gene ZINB parameters using three linear heads. The mean $M$, zero-inflation probability $\Pi$, and dispersion $\Theta$ are
\begin{equation}
\begin{aligned}
M &= \exp\left(\operatorname{clip}(ZW_{\mu}+b_{\mu},-15,15)\right),\\
\Pi &= \operatorname{sigmoid}(ZW_{\pi}+b_{\pi}),\\
\Theta &= \exp\left(\operatorname{clip}(ZW_{\theta}+b_{\theta},-15,15)\right).
\end{aligned}
\end{equation}
Moving these gene-dimensional projections after graph propagation avoids constructing gene-dimensional messages for every graph edge.

For an observation $x\geq0$, negative-binomial mean $\mu>0$, dispersion $\theta>0$, and zero-inflation probability $\pi\in(0,1)$, the negative-binomial log probability is
\begin{equation}
\begin{aligned}
\log p_{NB}(x\mid\mu,\theta)
={}&\log\Gamma(x+\theta)-\log\Gamma(\theta)\\
&-\log\Gamma(x+1)\\
&+\theta\left[\log\theta-\log(\theta+\mu)\right]\\
&+x\left[\log\mu-\log(\theta+\mu)\right].
\end{aligned}
\end{equation}
The ZINB log probability is
\begin{equation}
\log p_{ZINB}(x)=
\begin{cases}
\log\!\left[\pi+(1-\pi)p_{NB}(0\mid\mu,\theta)\right], & x=0,\\[2pt]
\log(1-\pi)+\log p_{NB}(x\mid\mu,\theta), & x>0.
\end{cases}
\end{equation}
The ZINB loss is the negative sum over cells and genes,
\begin{equation}
L_{ZINB}=-\sum_{i=1}^{n}\sum_{j=1}^{m}\log p_{ZINB}(X_{ij}).
\label{eq:zinb}
\end{equation}

\subsection{Expression Decoder and Joint Objective}
A separate decoder reconstructs expression directly from the latent representation. It contains a 1024-unit hidden layer with BatchNorm, ReLU, and dropout of 0.4, followed by a linear layer back to the original gene dimension and a ReLU activation:
\begin{equation}
X_D=\operatorname{ReLU}\left(W_2\operatorname{Dropout}_{0.4}\left(\operatorname{ReLU}(\operatorname{BN}(W_1Z+b_1))\right)+b_2\right).
\end{equation}
The decoded component is augmented by two normalized views of the input,
\begin{equation}
\widehat{X}=X_D+\operatorname{BN}_{g}(X)+\left[\operatorname{BN}_{c}(X^T)\right]^T.
\end{equation}
The reconstruction loss is mean squared error,
\begin{equation}
L_{MSE}=\frac{1}{nm}\sum_{i=1}^{n}\sum_{j=1}^{m}(X_{ij}-\widehat{X}_{ij})^2.
\end{equation}
The complete variational objective is
\begin{equation}
L=\alpha L_{ZINB}+(1-\alpha)L_{MSE}+\beta L_{KL},
\label{eq:total}
\end{equation}
where $\alpha=0.05$ and $\beta=10^{-4}$. The model is trained for 100 epochs using Adam~\cite{kingma2014adam} with learning rate $10^{-4}$. After training, $\widehat{X}$ is returned as the imputed expression matrix.

\section{Experiments}
\subsection{Datasets}
We evaluate scVGAE on 14 real-world scRNA-seq datasets spanning multiple tissues, species, sample sizes, and cell-type compositions. The datasets vary substantially in both the number of cells and the number of reference cell classes, providing a heterogeneous benchmark for evaluating downstream clustering performance. Dataset characteristics before preprocessing are summarized in Table~\ref{tab:data}.

\begin{table}[!t]
\caption{Characteristics of the 14 scRNA-seq benchmark datasets before
preprocessing. The number of reference cell classes, cells, and measured
genes used in the evaluation are reported.}
\label{tab:data}
\centering
\footnotesize
\renewcommand{\arraystretch}{1.05}
\setlength{\tabcolsep}{4pt}
\begin{tabular}{lrrr}
\toprule
Dataset & Classes & Cells & Genes \\
\midrule
Baron~\cite{baron_single-cell_2016} & 14 & 1937 & 20125 \\
Brosens~\cite{rawlings2021modelling} & 11 & 5829 & 20697 \\
Carey~\cite{strieder2022single} & 10 & 9082 & 27202 \\
cbmc~\cite{linderman2022zero} & 15 & 8617 & 20501 \\
Chang~\cite{boland2020heterogeneity} & 16 & 22770 & 10160 \\
Fujii~\cite{fujii_human_2018} & 9 & 2484 & 14679 \\
hcabm40k~\cite{hao2021integrated} & 8 & 12000 & 17369 \\
Hrvatin~\cite{hrvatin2018single} & 8 & 7390 & 19155 \\
Jakel~\cite{jakel2019altered} & 8 & 8000 & 21581 \\
Jiang~\cite{wang2022single} & 8 & 1014 & 25742 \\
Manno~\cite{la_manno_molecular_2016} & 26 & 1977 & 19531 \\
Mingyao~\cite{lakkis_multi-use_2022} & 8 & 6840 & 13007 \\
pbmc3k~\cite{hao2021integrated} & 9 & 2638 & 13714 \\
Xu~\cite{wang2022single} & 14 & 8196 & 19988 \\
\bottomrule
\end{tabular}
\end{table}

\subsection{Baselines and Evaluation Protocol}
We compare scVGAE with the un-imputed expression matrix (Original) and five established imputation methods: MAGIC~\cite{van2018recovering}, ALRA~\cite{linderman2022zero}, DeepImpute~\cite{arisdakessian2019deepimpute}, DCA~\cite{eraslan2019single}, and GNNImpute~\cite{xu2021efficient}. These baselines represent complementary modeling strategies, including affinity-based diffusion, low-rank approximation, deep neural reconstruction, count-distribution modeling, and graph-based imputation.

For downstream evaluation, each expression matrix is clustered using $k$-means and spectral clustering. Spectral clustering is evaluated with cosine, linear, and polynomial affinity functions, resulting in four clustering configurations in total. The number of clusters is set to the corresponding number of reference cell classes, and the random seed is fixed to 42 for reproducibility.

Clustering performance is quantified using the Adjusted Rand Index (ARI) and Adjusted Mutual Information (AMI), which measure agreement between predicted clusters and reference cell labels while accounting for chance agreement. Following the benchmark protocol used in this study, the reported ARI is the maximum ARI across the four clustering configurations, while the reported AMI is independently selected as the maximum AMI across the same configurations. If a spectral clustering configuration fails numerically for a dataset, the remaining valid configurations are used. Because this protocol selects the best-performing clustering configuration for each metric, the absolute ARI and AMI values should be interpreted accordingly.

All experiments are implemented in PyTorch 2.5.1 with CUDA 12.4 and executed on an NVIDIA L40S GPU with 48~GB of memory. Unless otherwise specified, scVGAE uses a latent dimension of 64, a graph neighborhood size of $k=30$, and a PCA dimension of 50. The model is trained for 100 epochs with the optimization settings described in Section~III-E.

\begin{table*}[!t]
\caption{Clustering performance across 14 scRNA-seq datasets measured
by adjusted Rand index (ARI) and adjusted mutual information (AMI).
For each dataset and metric, the best result is shown in bold, the
second-best is underlined, and the third-best is marked with
$\dagger$. The Overall row reports the arithmetic mean across all
14 datasets.}
\label{tab:clustering}
\centering
\fontsize{7.4}{8.4}\selectfont
\renewcommand{\arraystretch}{1.04}
\setlength{\tabcolsep}{3.2pt}
\begin{tabular}{llccccccc}
\toprule
Metric & Dataset & scVGAE & Original & MAGIC & ALRA & DeepImpute & DCA & GNNImpute \\
\midrule
\multirow{15}{*}{ARI}
& Baron & \underline{0.7922} & 0.5606 & 0.7657$^{\dagger}$ & 0.4235 & \textbf{0.8623} & 0.6392 & 0.5165 \\
& Brosens & \textbf{0.5721} & \underline{0.5083} & 0.4685 & 0.4967$^{\dagger}$ & 0.1991 & 0.3775 & 0.3746 \\
& Carey & 0.6586 & 0.6598$^{\dagger}$ & \textbf{0.7437} & 0.6510 & 0.3750 & 0.4799 & \underline{0.7023} \\
& cbmc & \underline{0.6159} & 0.4358 & \textbf{0.6622} & 0.4699 & 0.3348 & 0.4432 & 0.5632$^{\dagger}$ \\
& Chang & \underline{0.1646} & \textbf{0.1747} & 0.1620$^{\dagger}$ & 0.1351 & 0.1217 & 0.1502 & 0.0956 \\
& Fujii & \underline{0.3843} & 0.3565$^{\dagger}$ & \textbf{0.4439} & 0.3462 & 0.1403 & 0.1978 & 0.3041 \\
& hcabm40k & 0.0355 & \underline{0.0371} & 0.0347 & \textbf{0.0895} & 0.0246 & 0.0352 & 0.0370$^{\dagger}$ \\
& Hrvatin & \textbf{0.7829} & 0.7119$^{\dagger}$ & \underline{0.7771} & 0.6021 & 0.4492 & 0.6681 & 0.6282 \\
& Jakel & \textbf{0.6028} & \underline{0.5043} & 0.4117 & 0.4419 & 0.3589 & 0.4634$^{\dagger}$ & 0.4562 \\
& Jiang & 0.2613 & \underline{0.3421} & 0.2871$^{\dagger}$ & \textbf{0.3435} & 0.1379 & 0.2612 & 0.2284 \\
& Manno & 0.2157$^{\dagger}$ & \textbf{0.2929} & 0.2123 & \underline{0.2858} & 0.1269 & 0.1726 & 0.1883 \\
& Mingyao & 0.1491 & \textbf{0.2303} & 0.1778 & \underline{0.2020} & 0.1629 & 0.1855$^{\dagger}$ & 0.1735 \\
& pbmc3k & \underline{0.6432} & 0.6381$^{\dagger}$ & \textbf{0.6501} & 0.6370 & 0.2911 & 0.5122 & 0.5583 \\
& Xu & \underline{0.6749} & \underline{0.6749} & \textbf{0.6995} & 0.6629$^{\dagger}$ & 0.3445 & 0.4711 & 0.6531 \\
& \textbf{Overall} & \textbf{0.4681} & 0.4377$^{\dagger}$ & \underline{0.4640} & 0.4134 & 0.2807 & 0.3612 & 0.3914 \\
\midrule
\multirow{15}{*}{AMI}
& Baron & \underline{0.8285} & 0.7500 & \textbf{0.8356} & 0.7143 & 0.7691$^{\dagger}$ & 0.7567 & 0.6641 \\
& Brosens & \textbf{0.6636} & 0.6355$^{\dagger}$ & \underline{0.6494} & 0.6296 & 0.3557 & 0.5506 & 0.5434 \\
& Carey & \underline{0.7481} & 0.7213 & \textbf{0.7782} & 0.7088 & 0.5577 & 0.6164 & 0.7266$^{\dagger}$ \\
& cbmc & \underline{0.6800} & 0.6367 & \textbf{0.7462} & 0.6437$^{\dagger}$ & 0.5244 & 0.6156 & 0.6258 \\
& Chang & \underline{0.3062} & \textbf{0.3135} & 0.3023$^{\dagger}$ & 0.2558 & 0.2266 & 0.2855 & 0.1941 \\
& Fujii & \underline{0.5159} & 0.4767$^{\dagger}$ & \textbf{0.5658} & 0.4749 & 0.2289 & 0.3216 & 0.4276 \\
& hcabm40k & 0.0525$^{\dagger}$ & 0.0525$^{\dagger}$ & 0.0517 & \textbf{0.1311} & 0.0420 & \underline{0.0532} & 0.0498 \\
& Hrvatin & \underline{0.8517} & 0.8514$^{\dagger}$ & \textbf{0.8677} & 0.7979 & 0.6888 & 0.8155 & 0.6856 \\
& Jakel & \textbf{0.6557} & \underline{0.6316} & 0.5875 & 0.5986$^{\dagger}$ & 0.4643 & 0.5718 & 0.5742 \\
& Jiang & 0.4758 & \underline{0.5101} & 0.4964$^{\dagger}$ & \textbf{0.5360} & 0.2456 & 0.4017 & 0.3974 \\
& Manno & 0.4163 & \underline{0.4466} & 0.4301$^{\dagger}$ & \textbf{0.4574} & 0.2995 & 0.3779 & 0.3680 \\
& Mingyao & 0.2863 & \textbf{0.3224} & 0.2942 & \underline{0.3105} & 0.2555 & 0.3039$^{\dagger}$ & 0.2646 \\
& pbmc3k & \underline{0.7556} & 0.7248 & \textbf{0.7649} & 0.7389$^{\dagger}$ & 0.4693 & 0.6661 & 0.6615 \\
& Xu & \textbf{0.7846} & 0.7396 & \underline{0.7814} & 0.7414$^{\dagger}$ & 0.4942 & 0.5811 & 0.7011 \\
& \textbf{Overall} & \underline{0.5729} & 0.5581$^{\dagger}$ & \textbf{0.5822} & 0.5528 & 0.4015 & 0.4941 & 0.4917 \\
\bottomrule
\end{tabular}
\end{table*}

\subsection{Clustering Performance}
Table~\ref{tab:clustering} summarizes clustering performance across all 14 datasets using ARI and AMI. For each dataset and metric, the best result is shown in bold, the second-best is underlined, and the third-best is marked with $\dagger$. The Overall row reports the arithmetic mean across all datasets.

At the aggregate level, scVGAE achieves the highest mean ARI of 0.4681, narrowly exceeding MAGIC at 0.4640, and the second-highest mean AMI of 0.5729, behind MAGIC at 0.5822. These results indicate that scVGAE provides competitive clustering performance across heterogeneous scRNA-seq datasets rather than relying on strong performance in only a small subset of benchmarks. 

At the dataset level, scVGAE obtains the best ARI on Brosens, Hrvatin, and Jakel, and the best AMI on Brosens, Jakel, and Xu. It also ranks among the top three methods on 10 of the 14 datasets for ARI and 11 of the 14 datasets for AMI. Performance nevertheless varies substantially across datasets: scVGAE is particularly competitive on Baron, Brosens, cbmc, Hrvatin, Jakel, pbmc3k, and Xu, whereas lower scores are observed on hcabm40k, Jiang, Manno, and Mingyao. 

Taken together, the results suggest that the graph-aware variational representation preserves cell-class structure effectively across a broad range of datasets, although no single imputation strategy is uniformly optimal. The dataset-dependent behavior further motivates the ablation analysis in the following section, where we examine the contributions of the ZINB objective, direct expression reconstruction, and variational latent formulation.

\FloatBarrier
\subsection{Ablation Study}
We evaluate five model variants under the same 14-dataset protocol:
the full scVGAE model, removal of the KL-divergence term, removal of
the ZINB objective, removal of the MSE reconstruction objective, and
a deterministic variant that replaces stochastic sampling with the
posterior mean, $Z=\boldsymbol{\mu}_z$, while also removing KL
regularization. Table~\ref{tab:ablation} reports the mean ARI and AMI
across the 14 datasets.

\begin{table}[!ht]
\caption{Ablation study averaged across the 14 benchmark datasets.
Each variant removes one component of the full model or replaces
stochastic latent sampling with the posterior mean. Higher ARI and
AMI indicate better clustering performance.}
\label{tab:ablation}
\centering
\footnotesize
\renewcommand{\arraystretch}{1.05}
\setlength{\tabcolsep}{6pt}
\begin{tabular}{lcc}
\toprule
Variant & ARI & AMI \\
\midrule
Full scVGAE & \textbf{0.4694} & 0.5738 \\
w/o KL & 0.4683 & \textbf{0.5744} \\
w/o ZINB & 0.4465 & 0.5511 \\
w/o Reconstruction & 0.4624 & 0.5614 \\
Deterministic & 0.4541 & 0.5740 \\
\bottomrule
\end{tabular}
\end{table}

Removing the ZINB objective produces the largest overall degradation, reducing mean ARI from 0.4694 to 0.4465 and mean AMI from 0.5738 to 0.5511. This result indicates that distribution-aware modeling of sparse expression values provides the strongest contribution among the evaluated components. Removing the direct reconstruction objective also decreases both metrics, although the effect is smaller, suggesting that the ZINB and MSE objectives provide complementary supervision.

The variational components show a more nuanced effect. Replacing stochastic sampling with the posterior mean reduces mean ARI from 0.4694 to 0.4541, while AMI remains nearly unchanged. In contrast, removing the KL-divergence term alone changes performance only marginally, with ARI decreasing to 0.4683 and AMI slightly increasing to 0.5744. Given the relatively small KL weight ($\beta=10^{-4}$), these results suggest that the clustering benefit of the variational formulation is associated more strongly with stochastic latent sampling than with KL regularization alone.

Overall, the ablation results identify the ZINB objective as the most important component for clustering performance, with direct expression reconstruction providing an additional benefit. The variational latent formulation contributes primarily to ARI, whereas the isolated effect of KL regularization is limited under the present objective weighting.

\subsection{Limitations}
The present evaluation has several limitations. First, clustering-based metrics assess preservation of cell-class structure rather than direct recovery of unobserved transcript expression. Consequently, strong ARI or AMI does not necessarily imply accurate recovery of individual missing values. Second, the benchmark reports the best ARI and AMI across multiple clustering configurations. Although this protocol is applied consistently across all compared methods, selecting the best-performing configuration can increase absolute scores relative to evaluation with a single fixed clustering algorithm.

Third, performance varies substantially across datasets, indicating that imputation quality can depend on dataset-specific expression geometry and graph construction. The present study also fixes key hyperparameters, including the graph neighborhood size and latent dimensionality, rather than performing extensive dataset-specific optimization. Finally, the current evaluation does not directly assess the calibration or biological meaning of the uncertainty represented by the variational posterior.

Future work should therefore include controlled masking experiments for direct expression-recovery evaluation, sensitivity analyses of graph and latent-space hyperparameters, uncertainty calibration, and alternative graph-construction strategies.

\section{Conclusion}
We presented scVGAE, a variational graph autoencoder for single-cell RNA-seq imputation that combines graph-aware representation learning with distribution-aware expression reconstruction. Graph convolutional layers parameterize a Gaussian posterior for each cell, latent variables are obtained through stochastic reparameterization, and KL divergence regularizes the posterior toward an isotropic Gaussian prior. The resulting latent representation is decoded through both gene-wise ZINB parameter heads and a direct expression decoder, while a PCA-based 30-nearest-neighbor graph enables sparse and scalable message passing.

Across 14 real-world scRNA-seq datasets, scVGAE achieves the highest mean ARI and the second-highest mean AMI among the evaluated approaches. The ablation study further identifies the ZINB objective as the strongest contributor to clustering performance, with direct expression reconstruction providing an additional benefit. The variational components show a more nuanced effect: stochastic latent sampling is associated with improved ARI, whereas the isolated contribution of KL regularization is limited under the present objective weighting.

Overall, the results show that scVGAE provides competitive clustering performance across heterogeneous scRNA-seq datasets while producing an imputed expression matrix through a principled probabilistic graph representation. These findings support variational graph modeling as a promising framework for integrating cell-cell structure with distribution-aware single-cell expression reconstruction.

\bibliographystyle{IEEEtran}
\bibliography{sample,foundational}

@inproceedings{kipf2017semi,
  title={Semi-Supervised Classification with Graph Convolutional Networks},
  author={Kipf, Thomas N. and Welling, Max},
  booktitle={International Conference on Learning Representations},
  year={2017},
  eprint={1609.02907},
  archivePrefix={arXiv}
}

@misc{kipf2016variational,
  title={Variational Graph Auto-Encoders},
  author={Kipf, Thomas N. and Welling, Max},
  year={2016},
  eprint={1611.07308},
  archivePrefix={arXiv},
  note={NIPS Workshop on Bayesian Deep Learning}
}

@inproceedings{kingma2013auto,
  title={Auto-Encoding Variational Bayes},
  author={Kingma, Diederik P. and Welling, Max},
  booktitle={International Conference on Learning Representations},
  year={2014},
  eprint={1312.6114},
  archivePrefix={arXiv}
}

@inproceedings{cai2021graphnorm,
  title={GraphNorm: A Principled Approach to Accelerating Graph Neural Network Training},
  author={Cai, Tianle and Luo, Shengjie and Xu, Keyulu and He, Di and Liu, Tie-Yan and Wang, Liwei},
  booktitle={Proceedings of the 38th International Conference on Machine Learning},
  series={Proceedings of Machine Learning Research},
  volume={139},
  pages={1204--1215},
  year={2021}
}

@article{linderman2022zero,
  title={Zero-preserving imputation of single-cell RNA-seq data},
  author={Linderman, George C and Zhao, Jun and Roulis, Manolis and Bielecki, Piotr and Flavell, Richard A and Nadler, Boaz and Kluger, Yuval},
  journal={Nature Communications}, volume={13}, number={1}, pages={192}, year={2022}
}

@article {inoue2024bigcn,
	author = {Inoue, Yoshitaka and Kulman, Ethan and Kuang, Rui},
	title = {BiGCN: Leveraging Cell and Gene Similarities for Single-cell Transcriptome Imputation with Bi-Graph Convolutional Networks},
	elocation-id = {2024.04.05.588342},
	year = {2024},
	doi = {10.1101/2024.04.05.588342},
	publisher = {Cold Spring Harbor Laboratory},
	URL = {https://www.biorxiv.org/content/early/2024/04/10/2024.04.05.588342},
	eprint = {https://www.biorxiv.org/content/early/2024/04/10/2024.04.05.588342.full.pdf},
	journal = {bioRxiv}
}

@article{inoue2026drug,
  title={Drug Discovery in the Era of Artificial Intelligence: From Target Identification to Clinical Trials},
  author={Inoue, Yoshitaka and Hao, Nan and Lu, Yingzhou and Fu, Tianfan and Luna, Augustin},
  journal={Health Data Science},
  year={2026},
  publisher={AAAS}
}

@article{hao2021integrated,
  title={Integrated analysis of multimodal single-cell data},
  author={Hao, Yuhan and Hao, Stephanie and Andersen-Nissen, Erica and Mauck, William M and Zheng, Shiwei and Butler, Andrew and Lee, Maddie J and Wilk, Aaron J and Darby, Charlotte and Zager, Michael and others},
  journal={Cell}, volume={184}, number={13}, pages={3573--3587}, year={2021}
}

@article{qi2021sdimpute,
  title={SDImpute: a statistical block imputation method based on cell-level and gene-level information for dropouts in single-cell RNA-seq data},
  author={Qi, Jing and Zhou, Yang and Zhao, Zicen and Jin, Shuilin},
  journal={PLoS Computational Biology}, volume={17}, number={6}, pages={e1009118}, year={2021}
}

@article{tran2022novel,
  title={A novel method for single-cell data imputation using subspace regression},
  author={Tran, Duc and Tran, Bang and Nguyen, Hung and Nguyen, Tin},
  journal={Scientific Reports}, volume={12}, number={1}, pages={2697}, year={2022}
}

@article{van2018recovering,
  title={Recovering gene interactions from single-cell data using data diffusion},
  author={Van Dijk, David and Sharma, Roshan and Nainys, Juozas and Yim, Kristina and Kathail, Pooja and Carr, Ambrose J and Burdziak, Cassandra and Moon, Kevin R and Chaffer, Christine L and Pattabiraman, Diwakar and others},
  journal={Cell}, volume={174}, number={3}, pages={716--729}, year={2018}
}

@article{li2018accurate,
  title={An accurate and robust imputation method scImpute for single-cell RNA-seq data},
  author={Li, Wei Vivian and Li, Jingyi Jessica},
  journal={Nature Communications}, volume={9}, number={1}, pages={997}, year={2018}
}

@article{gong2018drimpute,
  title={DrImpute: imputing dropout events in single cell RNA sequencing data},
  author={Gong, Wuming and Kwak, Il-Youp and Pota, Pruthvi and Koyano-Nakagawa, Naoko and Garry, Daniel J},
  journal={BMC Bioinformatics}, volume={19}, pages={1--10}, year={2018}
}

@article{gu2022scgnn,
  title={scGNN 2.0: a graph neural network tool for imputation and clustering of single-cell RNA-Seq data},
  author={Gu, Haocheng and Cheng, Hao and Ma, Anjun and Li, Yang and Wang, Juexin and Xu, Dong and Ma, Qin},
  journal={Bioinformatics}, volume={38}, number={23}, pages={5322--5325}, year={2022}
}

@article{hrvatin2018single,
  title={Single-cell analysis of experience-dependent transcriptomic states in the mouse visual cortex},
  author={Hrvatin, Sinisa and Hochbaum, Daniel R and Nagy, M Aurel and Cicconet, Marcelo and Robertson, Keiramarie and Cheadle, Lucas and Zilionis, Rapolas and Ratner, Alex and Borges-Monroy, Rebeca and Klein, Allon M and others},
  journal={Nature Neuroscience}, volume={21}, number={1}, pages={120--129}, year={2018}
}

@article{arisdakessian2019deepimpute,
  title={DeepImpute: an accurate, fast, and scalable deep neural network method to impute single-cell RNA-seq data},
  author={Arisdakessian, C{\'e}dric and Poirion, Olivier and Yunits, Breck and Zhu, Xun and Garmire, Lana X},
  journal={Genome Biology}, volume={20}, number={1}, pages={1--14}, year={2019}
}

@article{eraslan2019single,
  title={Single-cell RNA-seq denoising using a deep count autoencoder},
  author={Eraslan, G{\"o}kcen and Simon, Lukas M and Mircea, Maria and Mueller, Nikola S and Theis, Fabian J},
  journal={Nature Communications}, volume={10}, number={1}, pages={390}, year={2019}
}

@article{xu2021efficient,
  title={An efficient scRNA-seq dropout imputation method using graph attention network},
  author={Xu, Chenyang and Cai, Lei and Gao, Jingyang},
  journal={BMC Bioinformatics}, volume={22}, pages={1--18}, year={2021}
}

@article{baron_single-cell_2016,
  title={A Single-Cell Transcriptomic Map of the Human and Mouse Pancreas Reveals Inter- and Intra-cell Population Structure},
  author={Baron, Maayan and Veres, Adrian and Wolock, Samuel L. and Faust, Aubrey L. and Gaujoux, Renaud and Vetere, Amedeo and Ryu, Jennifer Hyoje and Wagner, Bridget K. and Shen-Orr, Shai S. and Klein, Allon M. and Melton, Douglas A. and Yanai, Itai},
  journal={Cell Systems}, volume={3}, number={4}, pages={346--360.e4}, year={2016}, doi={10.1016/j.cels.2016.08.011}
}

@article{rawlings2021modelling,
  title={Modelling the impact of decidual senescence on embryo implantation in human endometrial assembloids},
  author={Rawlings, Thomas M and Makwana, Komal and Taylor, Deborah M and Mol{\`e}, Matteo A and Fishwick, Katherine J and Tryfonos, Maria and Odendaal, Joshua and Hawkes, Amelia and Zernicka-Goetz, Magdalena and Hartshorne, Geraldine M and others},
  journal={eLife}, volume={10}, pages={e69603}, year={2021}
}

@article{strieder2022single,
	author = {Strieder-Barboza, Clarissa and Flesher, Carmen G. and Geletka, Lynn M. and Delproposto, Jennifer B. and Eichler, Tad and Akinleye, Olukemi and Ky, Alexander and Ehlers, Anne P. and O{\textquoteright}Rourke, Robert W. and Lumeng, Carey N.},
	title = {Single-nuclei Transcriptome of Human AT Reveals Metabolically Distinct Depot-Specific Adipose Progenitor Subpopulations},
	elocation-id = {2022.06.29.496888},
	year = {2022},
	doi = {10.1101/2022.06.29.496888},
	publisher = {Cold Spring Harbor Laboratory},
	URL = {https://www.biorxiv.org/content/early/2022/06/29/2022.06.29.496888},
	eprint = {https://www.biorxiv.org/content/early/2022/06/29/2022.06.29.496888.full.pdf},
	journal = {bioRxiv}
}

@article{kingma2014adam,
  title={Adam: A method for stochastic optimization},
  author={Kingma, Diederik P and Ba, Jimmy}, journal={arXiv preprint arXiv:1412.6980}, year={2014}
}

@article{wang2022single,
  title={Single-cell analysis reveals a pathogenic cellular module associated with early allograft dysfunction after liver transplantation},
  author={Wang, Zheng and Shao, Xin and Wang, Kai and Lu, Xiaoyan and Zhuang, Li and Yang, Xinyu and Zhang, Ping and Yang, Penghui and Zheng, Shusen and Xu, Xiao and others},
  journal={bioRxiv}, year={2022}
}

@article{fujii_human_2018,
  title={Human Intestinal Organoids Maintain Self-Renewal Capacity and Cellular Diversity in Niche-Inspired Culture Condition},
  author={Fujii, Masayuki and Matano, Mami and Toshimitsu, Kohta and Takano, Ai and Mikami, Yohei and Nishikori, Shingo and Sugimoto, Shinya and Sato, Toshiro},
  journal={Cell Stem Cell}, volume={23}, number={6}, pages={787--793.e6}, year={2018}, doi={10.1016/j.stem.2018.11.016}
}

@article{lakkis_multi-use_2022,
  title={A multi-use deep learning method for CITE-seq and single-cell RNA-seq data integration with cell surface protein prediction and imputation},
  author={Lakkis, Justin and Schroeder, Amelia and Su, Kenong and Lee, Michelle Y. Y. and Bashore, Alexander C. and Reilly, Muredach P. and Li, Mingyao},
  journal={Nature Machine Intelligence}, volume={4}, number={11}, pages={940--952}, year={2022}, doi={10.1038/s42256-022-00545-w}
}

@article{jakel2019altered,
  title={Altered human oligodendrocyte heterogeneity in multiple sclerosis},
  author={J{\"a}kel, Sarah and Agirre, Eneritz and Mendanha Falc{\~a}o, Ana and Van Bruggen, David and Lee, Ka Wai and Knuesel, Irene and Malhotra, Dheeraj and Ffrench-Constant, Charles and Williams, Anna and Castelo-Branco, Gon{\c{c}}alo},
  journal={Nature}, volume={566}, number={7745}, pages={543--547}, year={2019}
}

@article{la_manno_molecular_2016,
  title={Molecular Diversity of Midbrain Development in Mouse, Human, and Stem Cells},
  author={La Manno, Gioele and Gyllborg, Daniel and Codeluppi, Simone and Nishimura, Kaneyasu and Salto, Carmen and Zeisel, Amit and Borm, Lars E. and Stott, Simon R. W. and Toledo, Enrique M. and Villaescusa, J. Carlos and L{\"o}nnerberg, Peter and Ryge, Jesper and Barker, Roger A. and Arenas, Ernest and Linnarsson, Sten},
  journal={Cell}, volume={167}, number={2}, pages={566--580.e19}, year={2016}, doi={10.1016/j.cell.2016.09.027}
}

@article{boland2020heterogeneity,
  title={Heterogeneity and clonal relationships of adaptive immune cells in ulcerative colitis revealed by single-cell analyses},
  author={Boland, Brigid S and He, Zhaoren and Tsai, Matthew S and Olvera, Jocelyn G and Omilusik, Kyla D and Duong, Han G and Kim, Eleanor S and Limary, Abigail E and Jin, Wenhao and Milner, J Justin and others},
  journal={Science Immunology}, volume={5}, number={50}, pages={eabb4432}, year={2020}
}

@article{luecken2019current,
  title={Current best practices in single-cell RNA-seq analysis: a tutorial},
  author={Luecken, Malte D and Theis, Fabian J}, journal={Molecular Systems Biology}, volume={15}, number={6}, pages={e8746}, year={2019}
}

@article{heath2016single,
  title={Single-cell analysis tools for drug discovery and development},
  author={Heath, James R and Ribas, Antoni and Mischel, Paul S}, journal={Nature Reviews Drug Discovery}, volume={15}, number={3}, pages={204--216}, year={2016}
}

@article{zhou2020graph,
  title={Graph neural networks: A review of methods and applications},
  author={Zhou, Jie and Cui, Ganqu and Hu, Shengding and Zhang, Zhengyan and Yang, Cheng and Liu, Zhiyuan and Wang, Lifeng and Li, Changcheng and Sun, Maosong},
  journal={AI Open}, volume={1}, pages={57--81}, year={2020}
}

@inproceedings{yu2022zinb,
  title={ZINB-based graph embedding autoencoder for single-cell RNA-seq interpretations},
  author={Yu, Zhuohan and Lu, Yifu and Wang, Yunhe and Tang, Fan and Wong, Ka-Chun and Li, Xiangtao},
  booktitle={Proceedings of the AAAI Conference on Artificial Intelligence}, volume={36}, number={4}, pages={4671--4679}, year={2022}
}

\end{document}